\documentclass[bn]{jjap3}
\usepackage{txfonts}

\title{
Spin polarization ratios of resistivity and density of states 
estimated from anisotropic magnetoresistance ratio 
for nearly half-metallic ferromagnets
}
\author{Satoshi Kokado$^1$\thanks{E-mail: 
kokado.satoshi@shizuoka.ac.jp
}, Yuya Sakuraba$^2$, 
and Masakiyo Tsunoda$^3$ 
}
\inst{
$^1$Graduate School of Integrated Science and Technology, 
Shizuoka University, Hamamatsu 432-8561, Japan \\
$^2$National Institute for Materials Science (NIMS), 
Tsukuba 305-0047, Japan \\
$^3$Department of Electronic Engineering, 
Graduate School of Engineering, Tohoku University, 
Sendai 980-8579, Japan

}
\abst{
We derive a simple relational expression between 
the spin polarization ratio of 
resistivity, $P_\rho$, 
and the anisotropic magnetoresistance ratio $\Delta \rho/\rho$, 
and that between 
the spin polarization ratio of the density of states at the Fermi energy, 
$P_{\rm DOS}$, and $\Delta \rho/\rho$ 
for nearly half-metallic ferromagnets. 
We find that 
$P_\rho$ and $P_{\rm DOS}$ increase 
with increasing $|\Delta \rho/\rho|$ from 0 to a maximum value. 
In addition, 
we roughly estimate $P_\rho$ and $P_{\rm DOS}$ 
for 
a Co$_2$FeGa$_{0.5}$Ge$_{0.5}$ Heusler alloy 
by 
substituting 
its experimentally observed $\Delta \rho/\rho$ 
into the respective expressions. 
}

\begin{document}
\maketitle

\begin{figure}
\begin{center}
\includegraphics[width=0.75\linewidth]{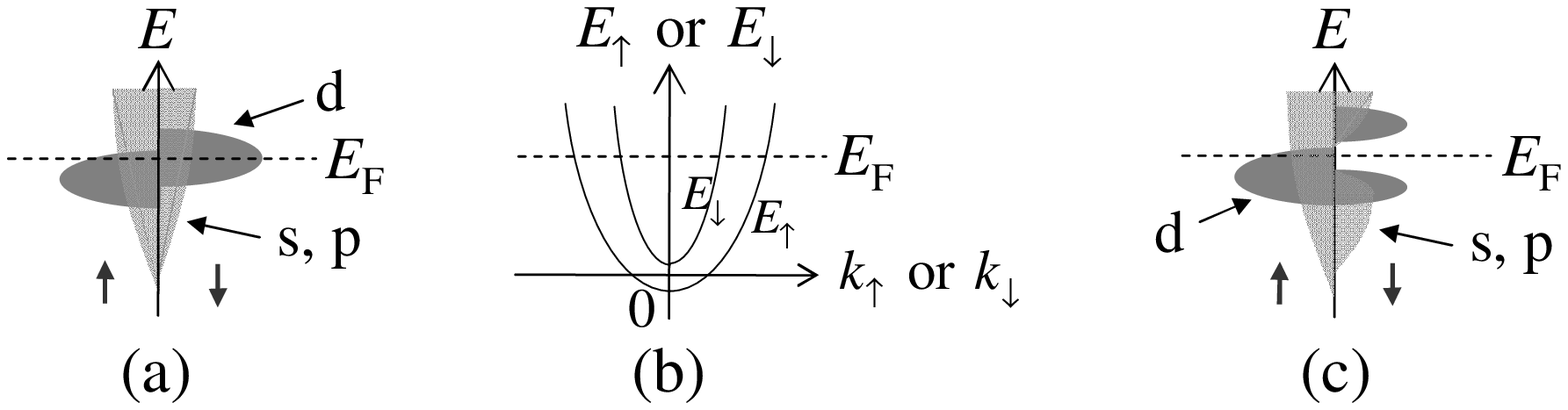}\\
\includegraphics[width=0.4\linewidth]{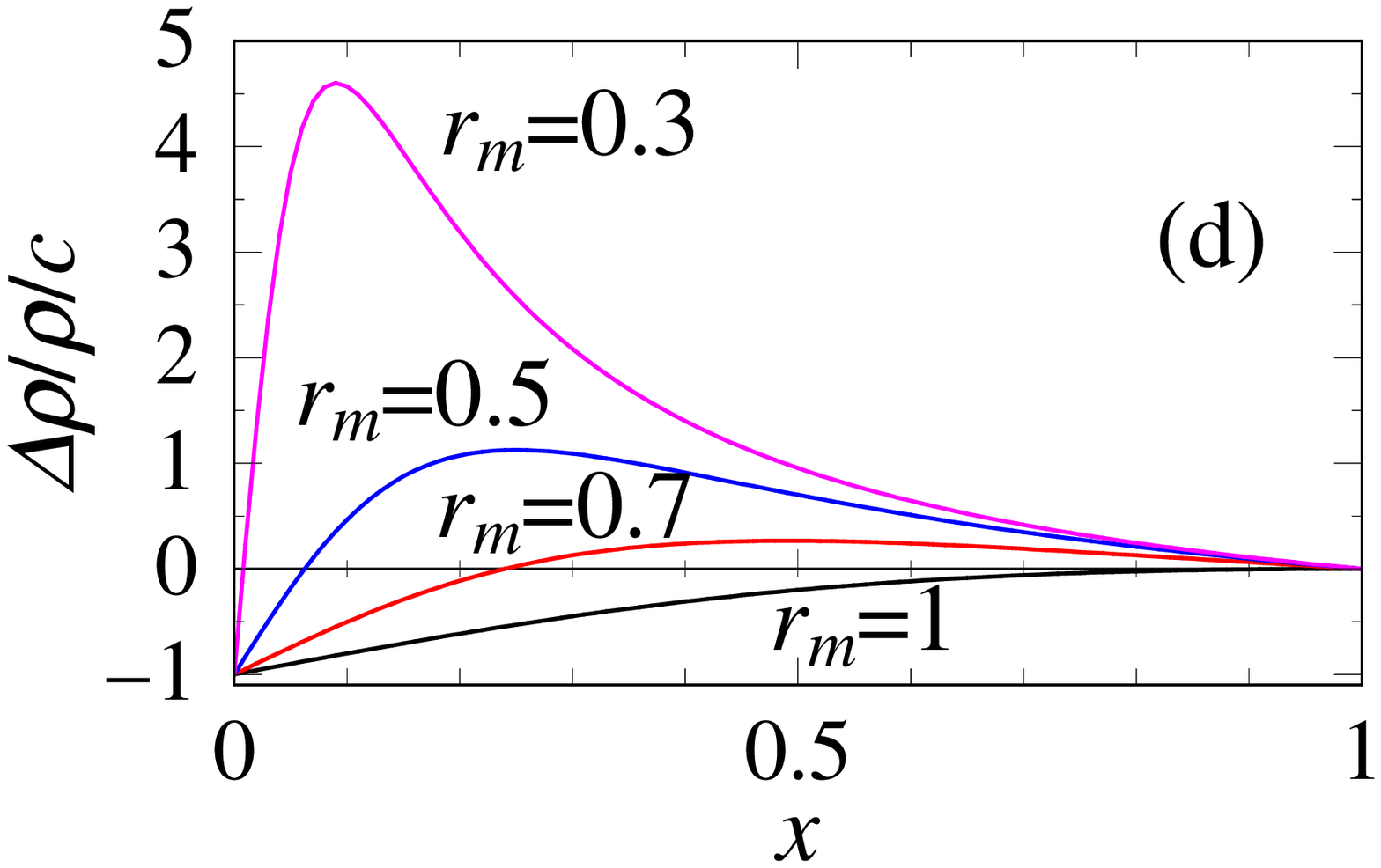}
\hspace{0.3cm}
\includegraphics[width=0.44\linewidth]{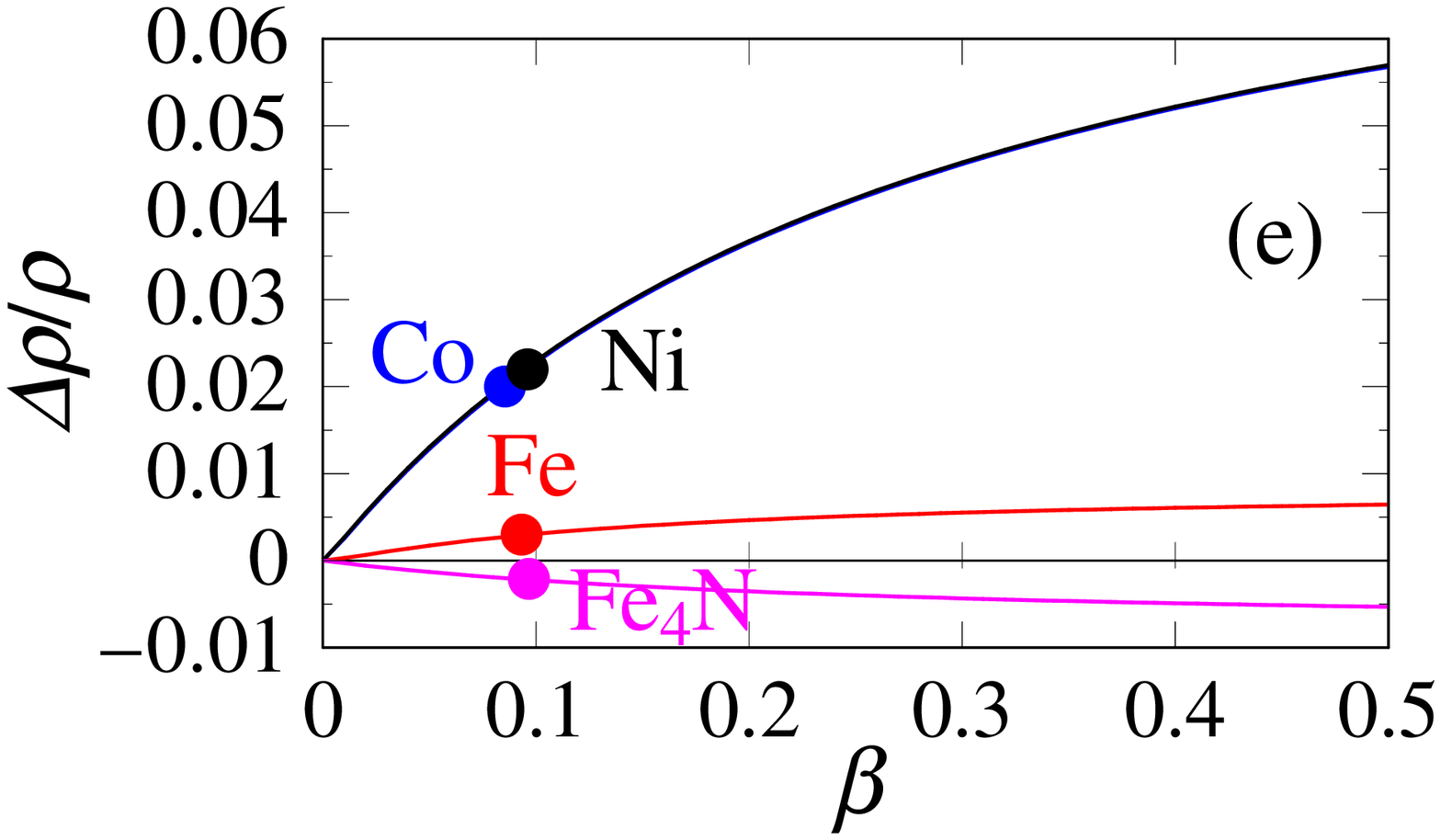}
\caption{
(Color online) 
(a) Partial DOSs of the s, p, and d states for the usual ferromagnets. 
(b) $E_\sigma$-$k_\sigma$ curve of the s and p states in (a). 
(c) Partial DOSs of the s, p, and d states 
for half-metallic Heusler alloys. 
In the case of the nearly HMF, 
there is a low DOS of the down spin at $E_{\mbox{\tiny F}}$. 
(d) 
$x$ dependence of $\Delta \rho/\rho/c$ of Eq. (\ref{AMR_x1}) 
with $x_s$=$x_d$$\equiv$$x$, 
$\beta_\uparrow$=$\beta_\downarrow$$\equiv$$\beta$, 
and $r_m$=0.3, 0.5, 0.7, and 1. 
(e) 
$\beta$ dependence of $\Delta \rho/\rho$ of Eq. (\ref{AMRx}) 
for Fe, Co, Ni, and Fe$_4$N is shown by solid curves. 
Here, 
we set 
$r_m$=1 
and use parameters in Table \ref{table}. 
The black, blue, red, and purple dots show 
experimental results of $\Delta \rho/\rho$ at 300 K 
for Ni, Co, Fe, and Fe$_4$N, respectively 
(see Table \ref{table}). 
\vspace{-0.5cm}
}
\label{amr}
\end{center}
\end{figure}

The anisotropic magnetoresistance (AMR) effect,
\cite{Thomson,McGuire,Miyazaki,Tsunoda,Yang,Sakuraba,Li1,Kokado1,Kokado2,Kokado3} 
in which the electrical resistivity depends on the magnetization direction, 
has been investigated using relatively easy experimental techniques 
for the last 160 years. 
The efficiency of the effect ``AMR ratio'' 
is generally defined by
\begin{eqnarray}
\label{AMR}
\Delta \rho/\rho=(\rho_\parallel - \rho_\perp)/\rho_\perp, 
\end{eqnarray}
where $\rho_\parallel$ ($\rho_\perp$) 
is 
the resistivity in the case of the electrical current 
parallel (perpendicular) to the magnetization. 
We recently derived the general expression of $\Delta \rho/\rho$ 
and found that 
$\Delta \rho/\rho$$<$0 is a necessary condition 
for a half-metallic ferromagnet (HMF)\cite{Kokado1,Kokado2}. 
The HMF is defined as having a finite density 
of states (DOS) at the Fermi energy $E_{\mbox{\tiny F}}$ 
in one spin channel 
and zero DOS at $E_{\mbox{\tiny F}}$ in the other spin channel 
[see Fig. \ref{amr}(c)]. 
Namely, 
the magnitude of the spin polarization ratio of the DOS 
at $E_{\mbox{\tiny F}}$, 
$|P_{\rm DOS}|$, is 1, 
where $P_{\rm DOS}$ is 
\begin{eqnarray}
\label{P_DOS}
P_{\rm DOS} = (D_\uparrow - D_\downarrow)/(D_\uparrow + D_\downarrow),
\end{eqnarray}
with $D_\uparrow$ ($D_\downarrow$) 
being the DOS of the up spin (down spin) at $E_{\mbox{\tiny F}}$. 
The above condition has been experimentally verified 
for Heusler alloys.\cite{Yang,Sakuraba} 

On the other hand, 
in recent years, 
a current-perpendicular-to-plane giant magnetoresistance (CPP-GMR) effect for 
ferromagnet/nonmagnetic-metal/ferromagnet pseudo spin valves 
has been actively studied 
for application to 
read sensors of future ultrahigh-density magnetic recording. 
In particular, studies to enhance the magnitude of the GMR effect 
are being carried out intensively. 
Here, the magnitude of this effect is represented by 
the resistance change area product $\Delta R A$, 
with $\Delta R$=$R_P - R_{AP}$, 
where $R_P$ ($R_{AP}$) is 
the resistance of the parallel (antiparallel) magnetization 
and $A$ is the area of the sample. 
According to the CPP-GMR theory by Valet and Fert\cite{Valet}, 
$\Delta R A$ is expressed by 
the spin polarization ratio of the resistivity of ferromagnets 
(the so-called bulk spin asymmetry coefficient), 
$P_\rho$, and so on. 
Here, $P_\rho$ is defined as 
\begin{eqnarray}
\label{P_rho}
P_\rho = (\rho_\downarrow - \rho_\uparrow )/(\rho_\uparrow + \rho_\downarrow),
\end{eqnarray}
where 
$\rho_\uparrow$ ($\rho_\downarrow$) 
is the resistivity of the up spin (down spin) of ferromagnets. 
The increase in $P_\rho$ tends to increase $\Delta R A$.\cite{Goripati,Li} 
For example, when ferromagnets are Heusler alloys, 
$\Delta R A$ 
becomes relatively large.\cite{Goripati} 

Recently, 
Sakuraba {\it et al}.\cite{Sakuraba} have experimentally observed 
the positive correlation between 
$|\Delta \rho/\rho|$ of 
the Co$_2$FeGa$_{0.5}$Ge$_{0.5}$ (CFGG) Heusler alloy\cite{BSDChS} 
and $\Delta R A$ of CFGG/Ag/CFGG pseudo spin valves. 
Here, this CFGG was regarded as a nearly HMF, 
in which there is a low DOS of the down spin at $E_{\mbox{\tiny F}}$ 
[see Fig. \ref{amr}(c)]. 
The correlation 
was considered on the basis of 
the relation 
between $\Delta \rho/\rho$ and $P_\rho$ 
mediated by $D_\downarrow/D_\uparrow$.\cite{Sakuraba} 
A relational expression between $\Delta \rho/\rho$ and $P_\rho$ 
and that between $\Delta \rho/\rho$ and $P_{\rm DOS}$, 
however, 
have scarcely been derived. 
Such expressions may make it possible to 
estimate $P_\rho$ and $P_{\rm DOS}$ from 
the relatively easy AMR measurements. 

In this paper, 
we derived 
a simple relational expression between 
$P_\rho$ and $\Delta \rho/\rho$ and 
that between 
$P_{\rm DOS}$ and $\Delta \rho/\rho$ 
for nearly HMFs 
using the two-current model. 
We found that 
$P_\rho$ and $P_{\rm DOS}$ increased 
with increasing $|\Delta \rho/\rho|$. 
We also estimated $P_\rho$ and $P_{\rm DOS}$ 
for CFGG 
by substituting 
its experimentally observed $\Delta \rho/\rho$ 
into the respective expressions.

We first report the general expression of $\Delta \rho/\rho$, 
which was previously derived by using the two-current model 
with the $s$-$s$ and $s$-$d$ scatterings.\cite{Kokado1,Kokado2} 
Here, 
$s$ denotes the conduction state of s, p, and conductive d states, 
and $d$ represents localized d states.\cite{Kokado1,Kokado2} 
The localized d states were obtained from a Hamiltonian with 
a spin--orbit interaction and an exchange field $H_{\rm ex}$. 
The AMR ratio $\Delta \rho/\rho$ was finally expressed as 
\begin{eqnarray}
\label{AMRx} 
\Delta \rho/\rho =-c (1-x_d) 
\left[ 1 - Z^2 \beta_\downarrow x_s/(\beta_\uparrow r_m^4) \right]\left/
(1 + Z x_s^2/r_m^4 ) \right., 
\end{eqnarray}
where 
$c$=$\gamma/[(\beta_\uparrow y_\uparrow)^{-1}+1]$ ($>$0), 
$Z$=$(1 + \beta_\uparrow y_\uparrow)/[1+(x_d/x_s) \beta_\downarrow y_\uparrow]$, 
$\gamma$=$(3/4) (\lambda/H_{\rm ex})^2$, 
$r_m$=$m_\downarrow^*/m_\uparrow^*$, 
$x_s$=$D_{s\downarrow}/D_{s\uparrow}$, 
$x_d$=$D_{d\downarrow}/D_{d\uparrow}$, 
$y_\uparrow$=$D_{d\uparrow}/D_{s\uparrow}$, 
$\beta_{\sigma}$=$n_{\rm imp}N_{\rm n} 
|V_{s\sigma \to d \sigma}|^2/
(n_{\rm imp}|V_s^{\rm imp}|^2 + |V_s^{\rm ph}|^2)$, 
and $\sigma$=$\uparrow$ or $\downarrow$. 
Here, 
$\lambda$ is the spin--orbit coupling constant, 
$n_{\rm imp}$ is the impurity density, 
$N_{\rm n}$ is the number of nearest-neighbor host atoms 
around the impurity, 
$V_{s\sigma \to d\sigma}$ is 
the matrix element 
for the $s$--$d$ scattering due to nonmagnetic impurities, 
$V_{s}^{\rm imp}$ is that 
for the $s$--$s$ scattering due to the impurities, 
and 
$V_{s}^{\rm ph}$ is that 
for the $s$--$s$ scattering due to phonons.\cite{Nishimura} 
The quantity $D_{s\sigma}$ is 
the partial DOS 
of the conduction state of the $\sigma$ spin at $E_{\mbox{\tiny F}}$ 
and 
$D_{d\varsigma}$ ($\varsigma$=$\uparrow$ or $\downarrow$) 
is the partial DOS 
of the localized d state of the magnetic quantum number $M$ 
and the $\varsigma$ spin at $E_{\mbox{\tiny F}}$ 
[see Fig. \ref{amr}(a)].\cite{Kokado1} 
In addition, 
$m_\sigma^*$ is an effective mass of 
electrons in the conduction band of the $\sigma$ spin, 
which is expressed as 
$\hbar^2 (d^2 E_\sigma /d k_\sigma^2)^{-1}$, 
where 
$E_\sigma$ is the energy of the conduction state of the $\sigma$ spin, 
$k_\sigma$ is the wave vector of the $\sigma$ spin in the current direction 
[see Fig. \ref{amr}(b)], 
and $\hbar$ is the Planck constant $h$ divided by 2$\pi$.\cite{Kittel} 
Note that 
Eq. (\ref{AMRx}) was derived under the assumption that 
the $s$--$s$ scattering rate 
is proportional to $D_{s\sigma}$ 
(i.e., the magnitude of 
the Fermi wave vector of the $\sigma$ spin).\cite{Kokado1} 
In the metallic case of Fig. \ref{amr}(a), 
therefore, 
Eq. (\ref{AMRx}) is effective 
at 0 K and in the temperature $T$ range of 
the $T$-linear resistivity 
including 300 K.\cite{Nishimura} 
On the other hand, in the HMF case of Fig. \ref{amr}(c), 
Eq. (\ref{AMRx}) is 
effective at 0 K and 
for $k_{\mbox{\tiny B}}T$$\ll$$E_c-E_{\mbox{\tiny F}}$ 
and $k_{\mbox{\tiny B}}T$$\ll$$E_{\mbox{\tiny F}}-E_v$, 
where $E_c$ ($E_v$) is the energy at the bottom of the conduction band 
(at the top of the valence band) 
of the down spin and 
$k_{\mbox{\tiny B}}$ is the Boltzmann constant. 
This restriction reflects that 
Eq. (\ref{AMRx}) does not take into account 
the thermal excitation of carriers. 

From Eq. (\ref{AMRx}), we next obtain a simple expression of 
$\Delta \rho/\rho$ 
with $x_s$=$x_d$$\equiv$$x$ 
to clearly show the effect of the DOS on $\Delta \rho/\rho$. 
Here, $x$ is assumed to be 0$\le$$x$$<$1, 
where $x$=0 ($x$$\ne$0) corresponds to the HMF 
(non-HMF) [see Fig. \ref{amr}(c)]. 
In addition, we set $\beta_\uparrow$=$\beta_\downarrow$$\equiv$$\beta$ 
for simplicity. 
Such simplifications permit 
only a rough estimation of $\Delta \rho/\rho$. 
Equation (\ref{AMRx}) then becomes 
\begin{eqnarray}
\label{AMR_x1}
\Delta \rho/\rho=-c(1-x)(r_m^4 - x)/(r_m^4 +x^2). 
\end{eqnarray}
Figure \ref{amr}(d) shows the $x$ dependence of $\Delta \rho/\rho$ 
of Eq. (\ref{AMR_x1}), 
with $r_m$=0.3, 0.5, 0.7, and 1. 
Each $\Delta \rho/\rho$ takes $-c$ at $x$=0 
and a positive maximum value at $x$=$r_m^2$ ($r_m$$\ne$1) 
and becomes closer to 0 as $x$ approaches 1. 
This behavior indicates that 
$\Delta \rho/\rho$$<$0 is the necessary condition for HMFs.

For Eq. (\ref{AMR_x1}), we now focus on 
nearly HMF cases with $\Delta \rho/\rho$$<$0;\cite{non-HMF} 
that is, 
$\Delta \rho/\rho$ is set to be $-|\Delta \rho/\rho|$. 
Utilizing Eq. (\ref{AMR_x1}) with 
$\Delta \rho/\rho$=$-|\Delta \rho/\rho|$, 
we derive 
the relational expression between $P_{\rm DOS}$ and $\Delta \rho/\rho$, 
and that between $P_\rho$ and $\Delta \rho/\rho$. 
The details are written as (i)--(iii):

(i) 
The quantity $x$ is obtained as solutions of Eq. (\ref{AMR_x1}), i.e., 
\begin{eqnarray}
\label{x_0}
&&x=0,~~{\rm for}~~|\Delta \rho/\rho|/c=1, \\
\label{x_pm1}
&&x = a - b~(\ne0),~~{\rm for}~~0<|\Delta \rho/\rho|/c<1, 
\end{eqnarray}
where 
$a$=$(1/2)(r_m^4 + 1)/(1 - |\Delta \rho/\rho|/c)$, 
$b$=$(1/2)\sqrt{d}$, 
and 
$d$=$[ (r_m^4+1)/(1 - |\Delta \rho/\rho|/c) ]^2 - 4 r_m^4$. 
Equations (\ref{x_0}) and (\ref{x_pm1}) 
correspond to 
the HMF and nearly HMF\cite{non-HMF} cases, respectively. 
As to Eq. (\ref{x_pm1}), 
we originally obtain $x_\pm$=$a \pm b$, 
where 0$<$$x_-$$<$1 and $x_+$$>$1. 
From the assumption of 0$\le$$x$$<$1, 
we choose $x_-$, i.e., Eq. (\ref{x_pm1}). 
The range 0$<$$|\Delta \rho/\rho|/c$$<$1 of Eq. (\ref{x_pm1}) 
is determined by considering 
$d$$\ge$0 
for 0$\le$$|\Delta \rho/\rho|/c$$\le$$1 + (r_m^4+1)/(2r_m^2)$ 
and 
$x$$>$0 
for 0$<$$|\Delta \rho/\rho|/c$$<$1, 
where 
$|a|$$>$$b$. 

(ii) The spin polarization ratio 
$P_{\rm DOS}$ of Eq. (\ref{P_DOS}) is written as 
\begin{eqnarray}
\label{P_DOS0}
P_{\rm DOS}= (1-x)/(1+x),
\end{eqnarray}
with 
$D_{\uparrow(\downarrow)}$=$D_{s \uparrow(\downarrow)} + \sum_{M=-2}^2 D_{d\uparrow(\downarrow)}$=$D_{s\uparrow(\downarrow)} + 5 D_{d\uparrow(\downarrow)}$ 
and $x_s$=$x_d$$\equiv$$x$. 
In the HFM case of Eq. (\ref{x_0}), 
$P_{\rm DOS}$ becomes 1. 
In the nearly HMF case of Eq. (\ref{x_pm1}), 
$P_{\rm DOS}$ is obtained by 
substituting $x$ of Eq. (\ref{x_pm1}) into Eq. (\ref{P_DOS0}):
\begin{eqnarray}
\label{P_DOS1}
&&
\hspace*{-0.8cm}
P_{\rm DOS}= \left[(r_m^4 +1) \left(2 - |\Delta \rho/\rho|/c \right) \right]^{-1} \Bigg[ \left(1- |\Delta \rho/\rho|/c \right)(1-r_m^4) 
+ \sqrt{ (r_m^4 + 1)^2 -4 r_m^4 \left( 1- |\Delta \rho/\rho|/c \right)^2} \Bigg]. \nonumber \\
\end{eqnarray}

(iii) 
The spin polarization ratio $P_\rho$ of Eq. (\ref{P_rho}) is obtained 
by using 
$\rho_\uparrow$=$\rho_{s\uparrow}+ \rho_{s\uparrow \to d \uparrow}$ 
and 
$\rho_\downarrow$=$\rho_{s\downarrow}+ \rho_{s\downarrow \to d \downarrow}$ 
in the two-current model,\cite{Kokado1} 
where 
$\rho_{s\sigma}$ ($\rho_{s\sigma \to d \varsigma}$) 
is the resistivity 
due to the $s$--$s$ scattering ($s$--$d$ scattering).\cite{Nishimura} 
In $\rho_\uparrow$ and $\rho_\downarrow$, 
terms with $\gamma$ are ignored 
because the effect of $\gamma$ on $P_\rho$ is negligibly small.\cite{gamma} 
As a result, 
$P_\rho$ is written as
\begin{eqnarray}
\label{P_rho0}
P_\rho=
\frac{
r_m^4 ( 1  + \beta_\downarrow y_\downarrow) 
- x_s^2 (1 + \beta_\uparrow y_\uparrow )}
{r_m^4 ( 1 + \beta_\downarrow y_\downarrow) 
+ x_s^2 (1 + \beta_\uparrow y_\uparrow )}, 
\end{eqnarray}
where $y_\downarrow$=$D_{d\downarrow}/D_{s\downarrow}$, 
$\rho_{s\downarrow}/\rho_{s\uparrow}$=$r_m^4/x_s^2$, 
$\rho_{s \sigma \to d \varsigma}/\rho_{s \sigma}$=$\beta_\sigma (D_{d\varsigma}/D_{s\sigma})$, 
and 
$\rho_{s \sigma' \to d \varsigma}/\rho_{s \sigma}$=$(\rho_{s \sigma'}/\rho_{s \sigma})\beta_{\sigma'} (D_{d\varsigma}/D_{s\sigma'})$ in Ref. \citen{Kokado1}, 
with $\sigma$, $\sigma'$, and $\varsigma$=$\uparrow$ or $\downarrow$. 
When $x_s$=$x_d$$\equiv$$x$ (i.e., $y_\uparrow$=$y_\downarrow$) 
and $\beta_\uparrow$=$\beta_\downarrow$$\equiv$$\beta$, 
Eq. (\ref{P_rho0}) is rewritten as
\begin{eqnarray}
\label{P_rho1}
P_\rho=(r_m^4-x^2)/(r_m^4+x^2). 
\end{eqnarray}
In the HMF case of Eq. (\ref{x_0}), 
$P_\rho$ 
becomes 1.\cite{thermal} 
In the nearly HMF case of Eq. (\ref{x_pm1}) (i.e., metallic case), 
$P_\rho$ is obtained by 
substituting $x$ of Eq. (\ref{x_pm1}) into Eq. (\ref{P_rho1}): 
\begin{eqnarray}
\label{P_rho2}
P_\rho=
(r_m^4 +1)^{-1}
\sqrt{(r_m^4 + 1)^2 -4 r_m^4 \left( 1 - |\Delta \rho/\rho|/c \right)^2 }. 
\end{eqnarray}

\begin{figure}
\begin{center}
\includegraphics[width=0.42\linewidth]{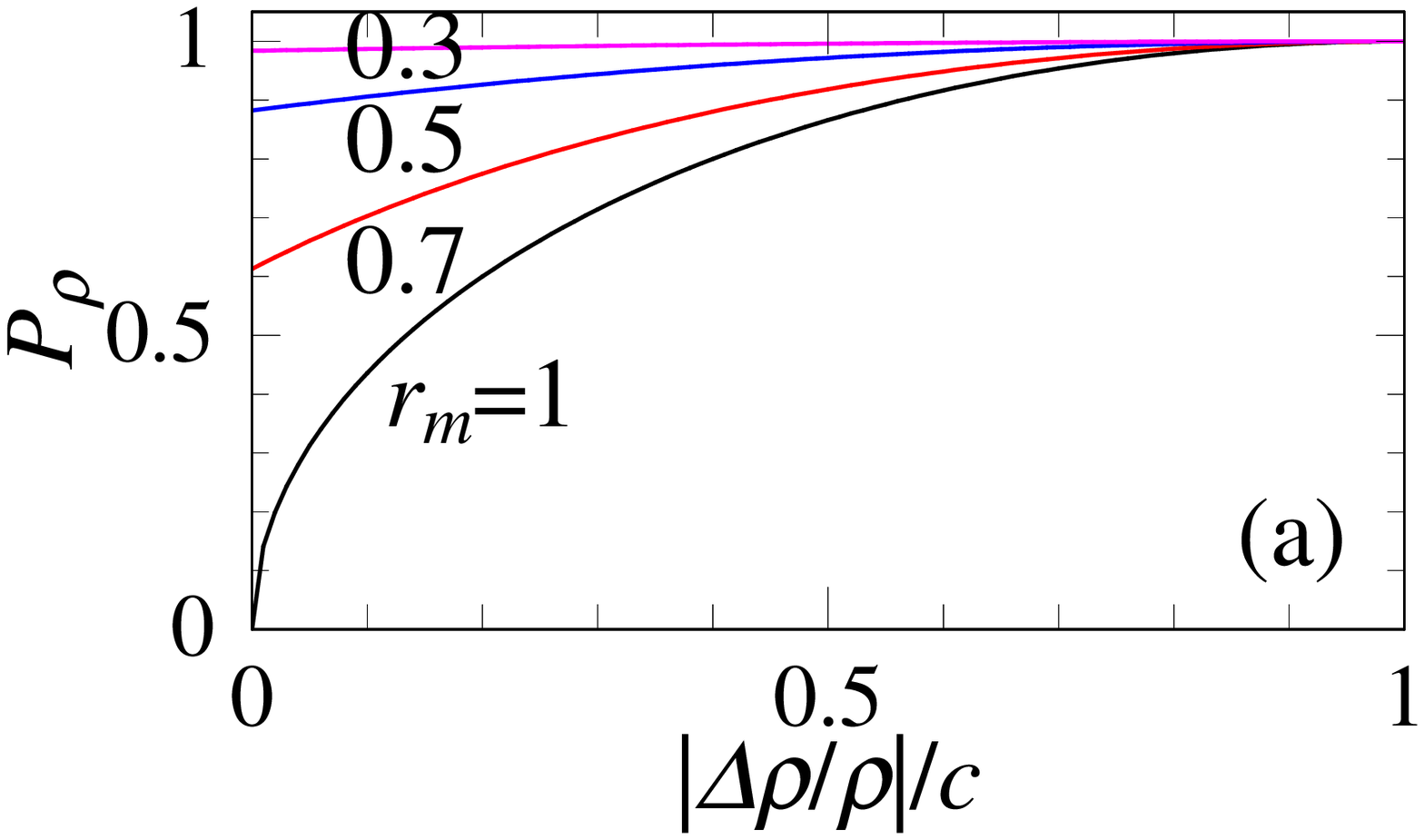}
\hspace{0.1cm}
\includegraphics[width=0.42\linewidth]{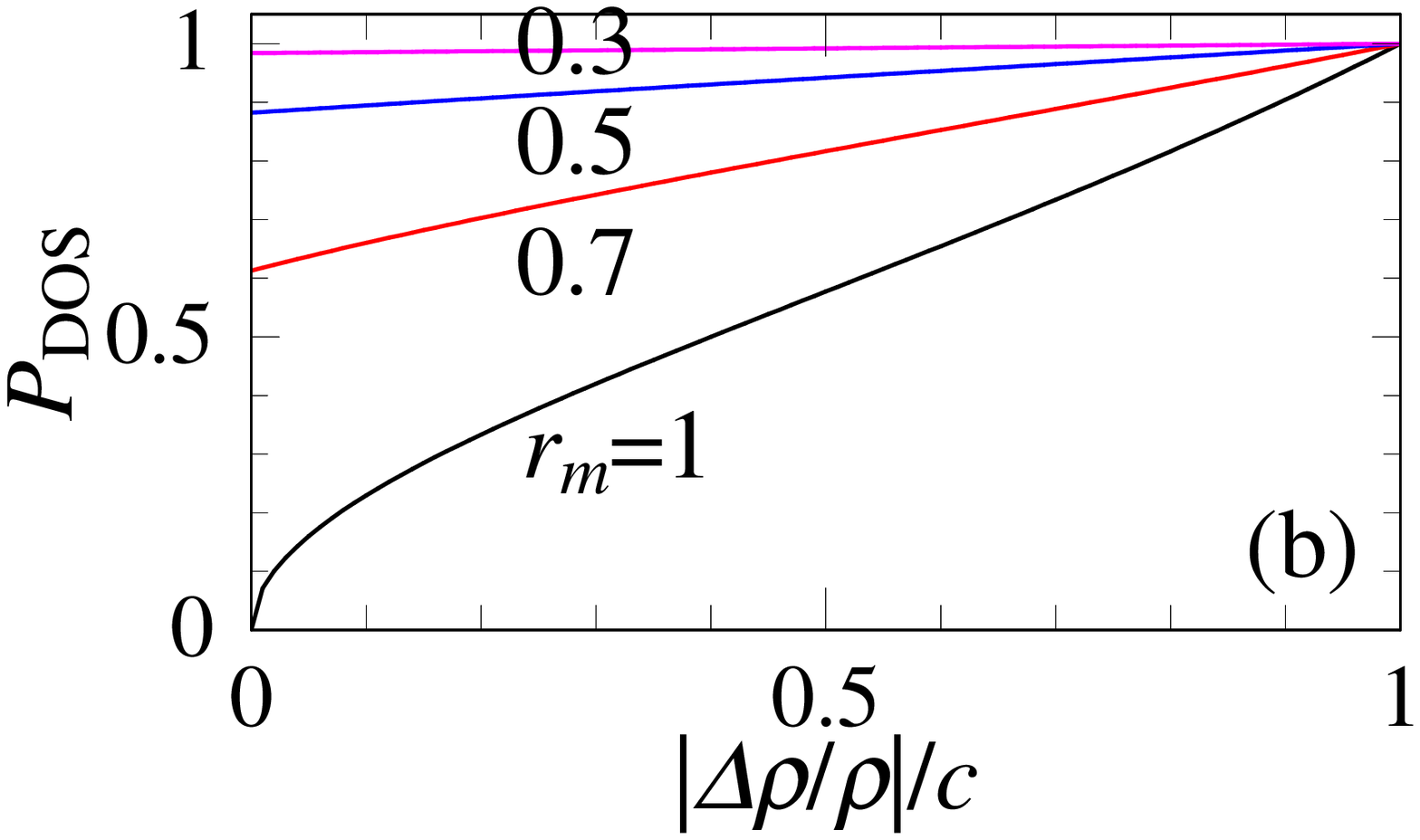}\\
\includegraphics[width=0.42\linewidth]{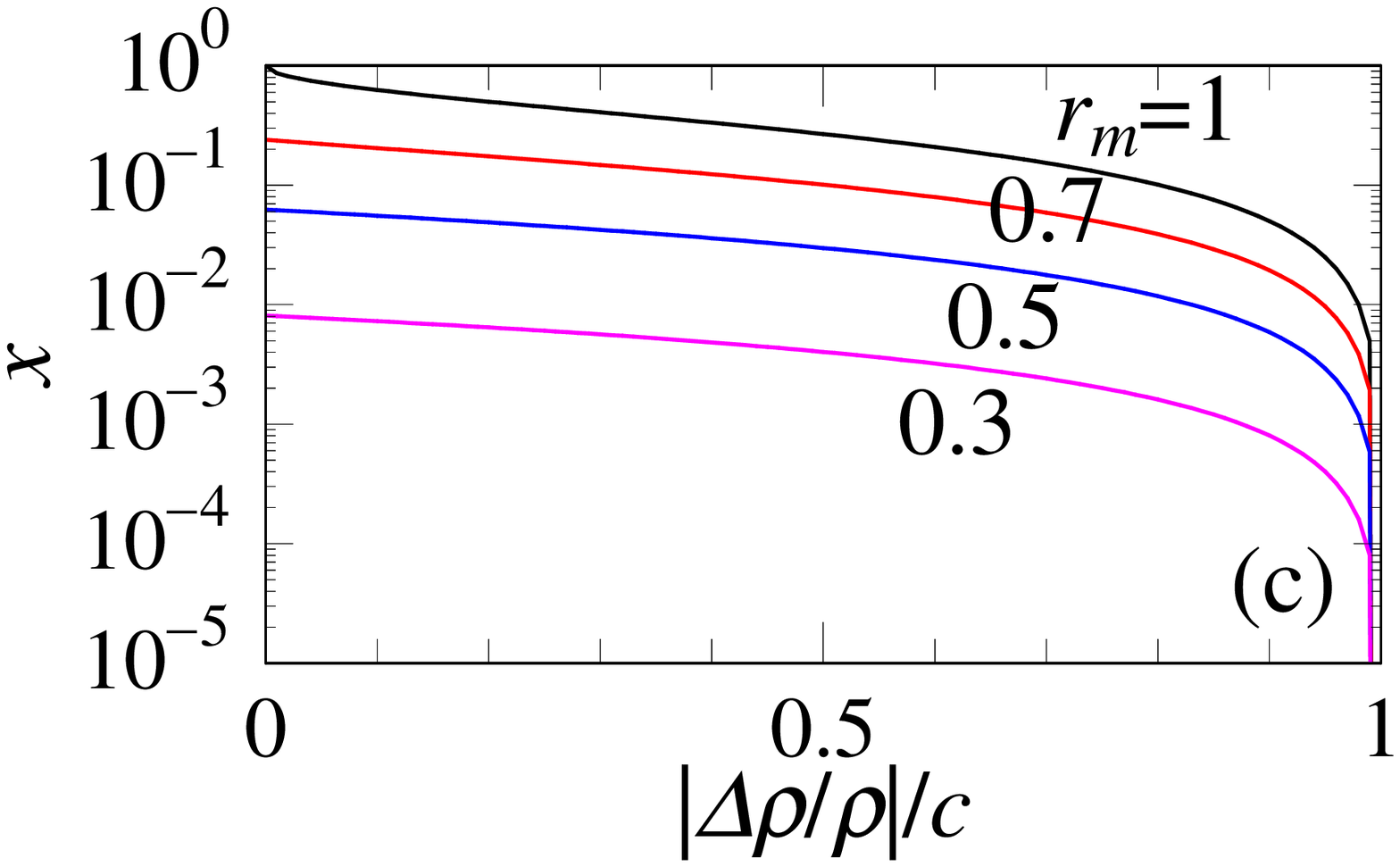}
\hspace{0.1cm}
\includegraphics[width=0.42\linewidth]{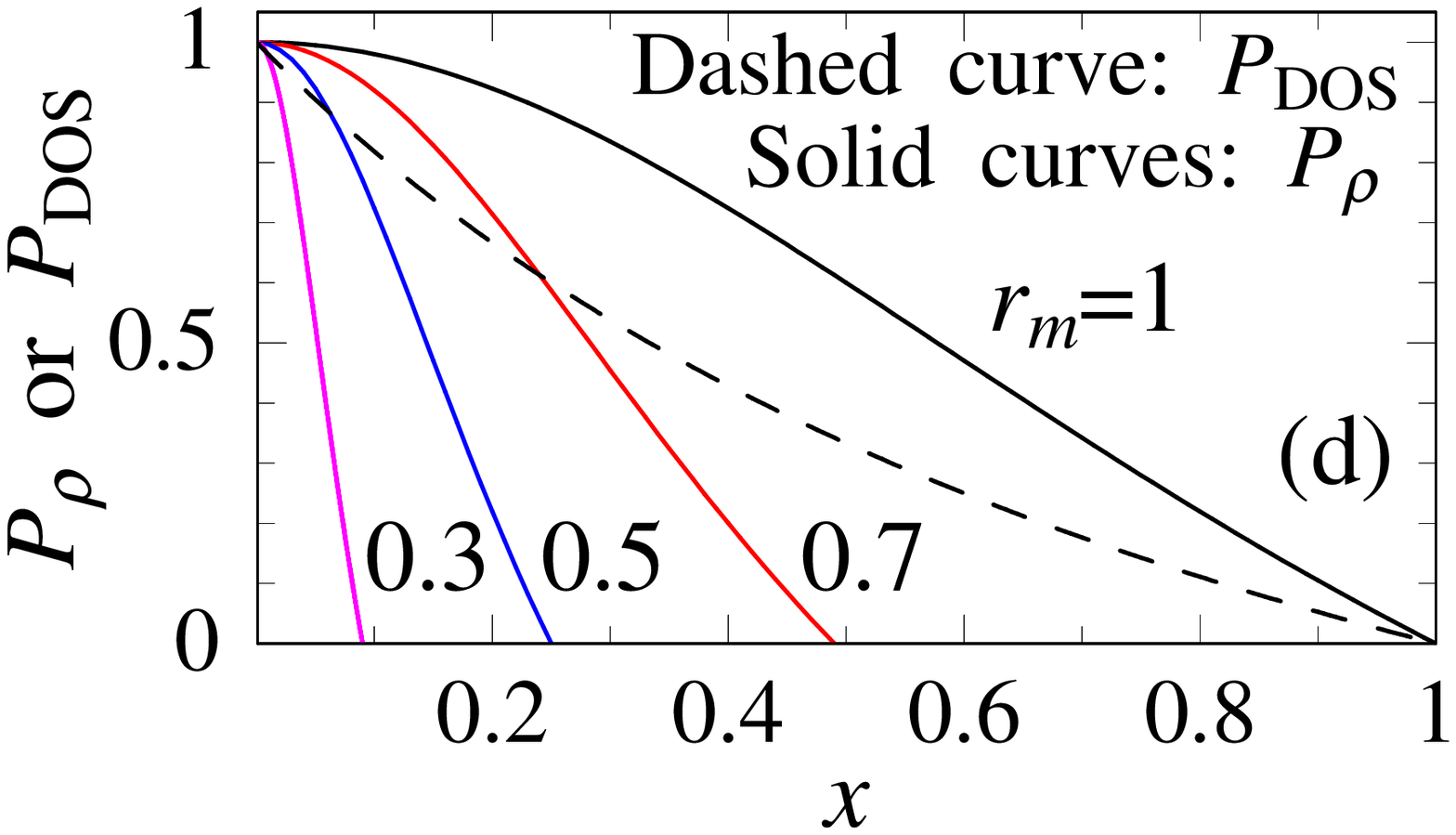}
\caption{
(Color online) 
(a) $|\Delta \rho/\rho|/c$ dependence of 
$P_\rho$ of Eq. (\ref{P_rho2}). 
(b) $|\Delta \rho/\rho|/c$ dependence of 
$P_{\rm DOS}$ of Eq. (\ref{P_DOS1}). 
(c) $|\Delta \rho/\rho|/c$ dependence of 
$x$ of Eq. (\ref{x_pm1}). 
(d) $x$ dependences of 
$P_\rho$ of Eq. (\ref{P_rho1}) 
and $P_{\rm DOS}$ of Eq. (\ref{P_DOS0}). 
Here, we set $r_m$=0.3, 0.5, 0.7, and 1, 
and $x_s$=$x_d$$\equiv$$x$. 
\vspace{-0.5cm}
}
\label{basic}
\end{center}
\end{figure}

In Figs. \ref{basic}(a) and \ref{basic}(b), 
we show the $|\Delta \rho/\rho|/c$ dependences of 
$P_\rho$ of Eq. (\ref{P_rho2}) 
and $P_{\rm DOS}$ of Eq. (\ref{P_DOS1}), respectively, 
where $r_m$=0.3, 0.5, 0.7, and 1. 
We find the positive correlation 
between $P_\rho$ and $|\Delta \rho/\rho|/c$, 
and that between $P_{\rm DOS}$ and $|\Delta \rho/\rho|/c$. 
Namely, 
$P_\rho$ and $P_{\rm DOS}$ increase to 1 
with increasing $|\Delta \rho/\rho|/c$ from 0 to 1 (maximum value). 
The reason for this is that 
the increase in $|\Delta \rho/\rho|/c$ decreases $x$ 
[see Fig. \ref{basic}(c)] 
and then 
the decrease in $x$ increases 
$P_\rho$ and $P_{\rm DOS}$ [see Fig. \ref{basic}(d)]. 
Furthermore, $P_\rho$ and $P_{\rm DOS}$ increase with decreasing $r_m$. 
The reason for this is that 
the decrease in $r_m$ reduces the maximum value of $x$ 
[see Fig. \ref{basic}(c)] 
and narrows the range of $x$, 
and then that feature of $x$ 
increases $P_\rho$ and $P_{\rm DOS}$ 
[see Fig. \ref{basic}(d)].

\begin{figure}[ht]
\begin{center}
\includegraphics[width=0.46\linewidth]{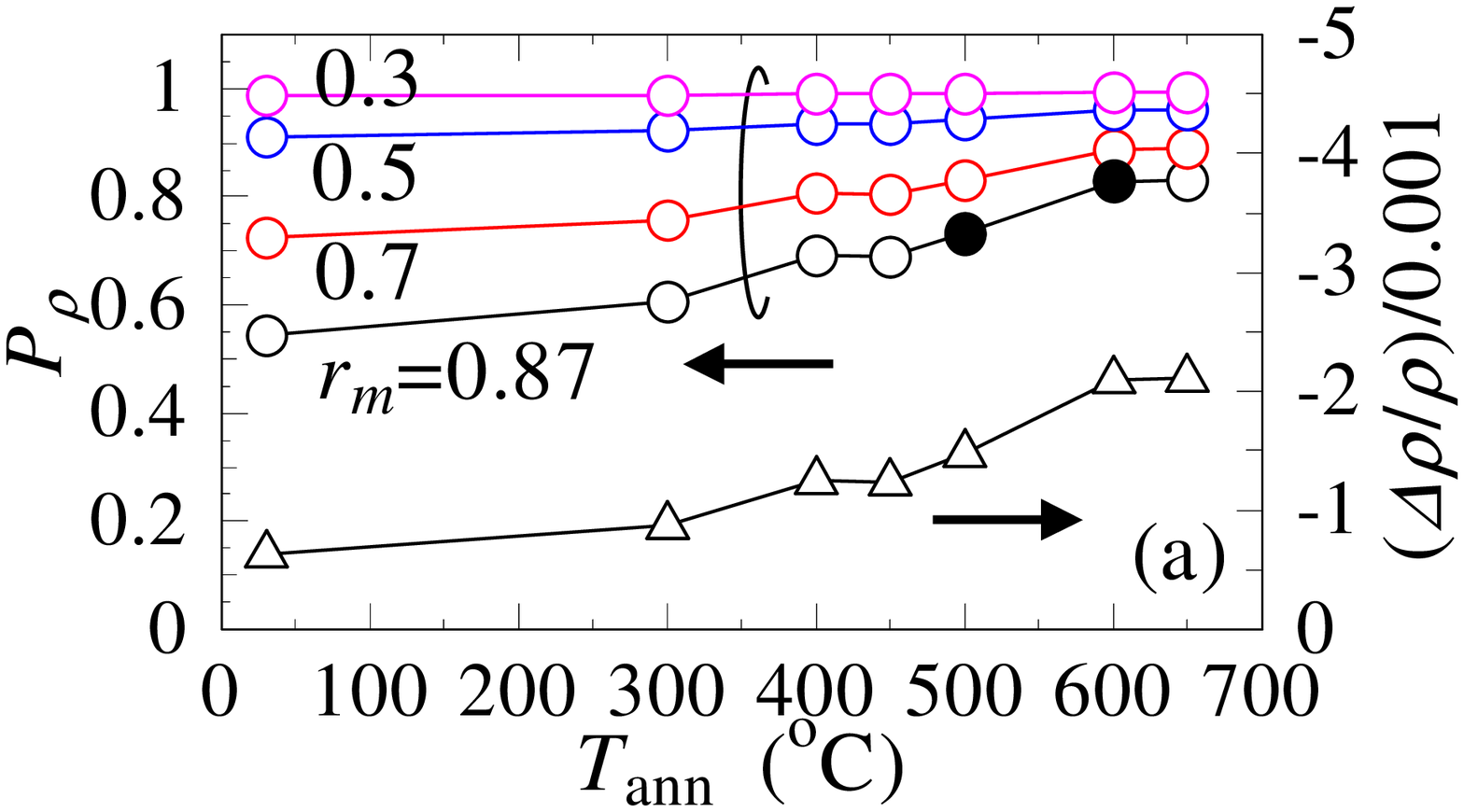}
\hspace{0.1cm}
\includegraphics[width=0.42\linewidth]{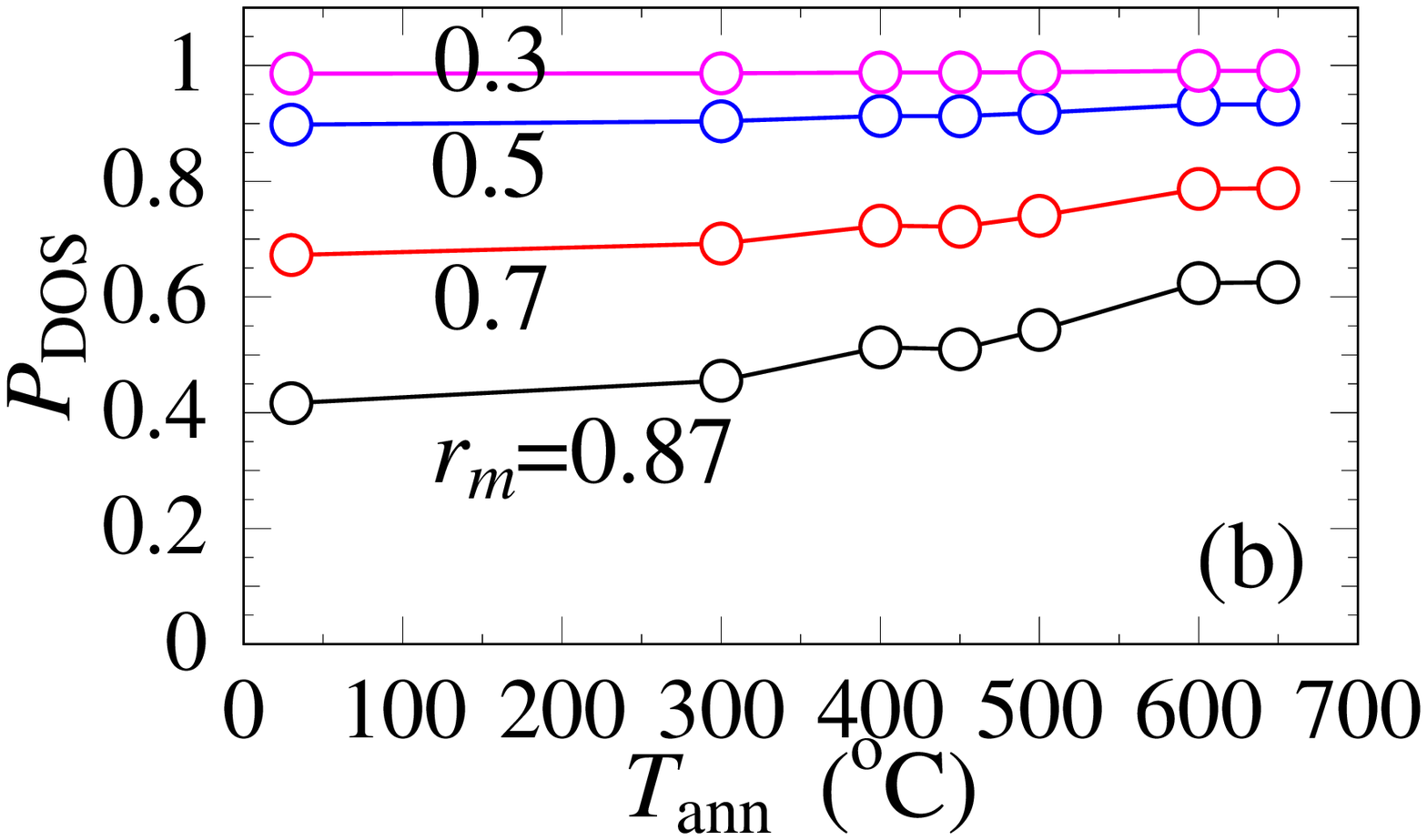}
\caption{
(Color online) 
(a) 
The white circles show 
the $T_{\rm ann}$ dependence of 
$P_\rho$ of Eq. (\ref{P_rho2}) for CFGG, where $r_m$=0.3, 0.5, 0.7, and 0.87. 
The respective black dots denote 
the previously evaluated $P_\rho$ 
at $T_{\rm ann}$=500\cite{Goripati} 
and 600 $^{\circ}$C\cite{Li} in Table \ref{table2}. 
The triangles show 
the experimental result of the $T_{\rm ann}$ dependence of 
$\Delta \rho/\rho$ at 300 K for CFGG.\cite{Sakuraba} 
(b) 
The $T_{\rm ann}$ dependence of 
$P_{\rm DOS}$ of Eq. (\ref{P_DOS1}) for CFGG 
with $r_m$=0.3, 0.5, 0.7, and 0.87. 
\vspace{-0.5cm}
}
\label{mater}
\end{center}
\end{figure}

As an application, 
we investigate $P_\rho$ and $P_{\rm DOS}$ for CFGG. 
Regarding parameters, 
we first set $\gamma$=0.01 as a typical value.\cite{Kokado1} 
The quantity $y_\uparrow$ is roughly estimated 
to be 10 
from the partial DOSs of similar Heusler alloys.\cite{Sharma}
Next, we consider the uncertain parameter 
$\beta$, 
which includes information on impurities and phonons. 
Although $\beta$ actually depends on materials, 
we determine $\beta$ 
from the $\beta$ dependence of $\Delta \rho/\rho$ 
for Fe, Co, Ni, and Fe$_4$N in Fig. \ref{amr}(e), 
where the respective parameters 
are noted in Table \ref{table}. 
By comparing the calculation results of Eq. (\ref{AMRx}) 
with the experimental results of $\Delta \rho/\rho$ 
at 300 K in Table \ref{table}, 
$\beta$ is roughly evaluated to be 0.1 [see Fig. \ref{amr}(e)]. 
This $\beta$=0.1 is used for the present systems. 
The constant $c$ is thus determined to be 0.005; 
that is, 
$|\Delta \rho/\rho|$ can take 
$c$=0.005 at 300 K for the HMF of $x$=0.\cite{condition} 
This $c$ increases with decreasing $T$ 
due to the decrease in $|V_s^{\rm ph}|^2$. 
Judging from the experimental result of 
$\Delta \rho/\rho$$\sim$$-$0.003 
at 10 K in Ref. \citen{Sakuraba} 
(i.e., $|\Delta \rho/\rho|$$<$0.005), 
the present CFGG appears to be a nearly HMF at 10 K. 
Under the assumption that 
the CFGG is a nearly HMF at 300 K as well as at 10 K, 
we roughly estimate 
the annealing temperature $T_{\rm ann}$ dependences of 
$P_\rho$ and $P_{\rm DOS}$ for CFGG 
by substituting its experimental result of $\Delta \rho/\rho$ at 300 K 
[see triangles in Fig. \ref{mater}(a)] 
into Eqs. (\ref{P_rho2}) and (\ref{P_DOS1}), respectively. 
The white circles in Figs. \ref{mater}(a) and \ref{mater}(b) 
indicate 
the $T_{\rm ann}$ dependences of $P_{\rho}$ and $P_{\rm DOS}$, 
respectively, where $r_m$=0.3, 0.5, 0.7, and 0.87.\cite{r_m<1} 
We find that 
$P_\rho$ and $P_{\rm DOS}$ increase 
with increasing $|\Delta \rho/\rho|$ 
and decreasing $r_m$ 
in the same trend as the results in Fig. \ref{basic}. 
Such $P_\rho$ is compared with the previous values 
at $T_{\rm ann}$=500 and 600 $^{\circ}$C 
in Table \ref{table2} [see black dots in Fig. \ref{mater}(a)], 
which were evaluated by fitting 
Valet--Fert's expression\cite{Valet} to the experimental results 
of the CFGG thickness dependence of 
$\Delta R A$ at 300 K.\cite{Goripati,Li} 
Since $P_{\rho}$ at $r_m$=0.87 
agrees with the previous values, 
we choose $r_m$=0.87 for the present system 
(see Fig. \ref{mater} and Table \ref{table2}).\cite{r_m,r_m<1} 
Table \ref{table2} also shows 
$P_{\rm DOS}$ ($\ne$1) at $r_m$=0.87. 
In general, $P_{\rm DOS}$$\ne$1 is considered to 
originate from atomic disorders,\cite{Li1} 
the decrease in $|H_{ex}|$,\cite{Kokado1} and so on. 
The origin of the present $P_{\rm DOS}$$\ne$1, however, 
has not yet been identified. 


In summary, 
we derived the simple relational expression between 
$P_\rho$ and $\Delta \rho/\rho$, 
and 
that between 
$P_{\rm DOS}$ and $\Delta \rho/\rho$ for nearly HMFs. 
In these expressions, 
$P_\rho$ and $P_{\rm DOS}$ increased to 1 
with increasing $|\Delta \rho/\rho|/c$ from 0 to 1 (maximum value). 
In addition, we roughly estimated 
$P_\rho$ and $P_{\rm DOS}$ for CFGG 
using the respective expressions.

%

\newpage
\noindent

\newpage
\noindent

\begin{table}
\caption{
Parameters $x_s$, $x_d$, and $y_\uparrow$, 
and experimental values of $\Delta \rho/\rho$ at 300 K 
for bcc Fe, fcc Co, fcc Ni, and Fe$_4$N. 
Each $x_s$ is evaluated from 
the values of $\rho_{s\downarrow}/\rho_{s\uparrow}$ and 
$\rho_{s\downarrow}/\rho_{s\uparrow}$=$r_m^4 (D_{s\uparrow}/D_{s\downarrow})^2$ (Ref. \citen{Kokado1}) with $r_m$=1. 
}
\begin{center}
\begin{tabular}{ccccc}
\hline 
Material & $x_s$ & $x_d$ \cite{Kokado1} & $y_\uparrow$ & $\Delta \rho/\rho$ (experiment)\\
\hline 
bcc Fe & 1.6 & 0.50 & 25 (Ref. \citen{papa}) & 0.0030 (Ref. \citen{McGuire}) \\
fcc Co & 0.37 & 10 & 3.5 (Ref. \citen{comment}) & 0.020 (Ref. \citen{McGuire}) \\
fcc Ni & 0.32 & 10 & 3.5 (Ref. \citen{papa}) & 0.022 (Ref. \citen{McGuire}) \\
Fe$_4$N & 25 & 5.0 & 20 (Ref. \citen{Sakuma}) & $-$0.0021 (Ref. \citen{Tsunoda}) \\
\hline 
\vspace{-0.8cm}
\end{tabular}
\end{center}
\label{table}
\end{table} 

\begin{table}
\caption{
Spin polarization ratios 
$P_{\rm DOS}$ of Eq. (\ref{P_DOS1}) and $P_\rho$ of Eq. (\ref{P_rho2}) 
at $T_{\rm ann}$=500 and 600 $^{\circ}$C for CFGG. 
They are 
the respective values at $r_m$=0.87 
in Figs. \ref{mater}(a) and \ref{mater}(b). 
The previous values of $P_\rho$, 
which were evaluated on the basis of $\Delta RA$ at 300 K,\cite{Goripati,Li} 
are also noted. 
}
\begin{center}
\begin{tabular}{cccc}
\hline 
$T_{\rm ann}$ ($^{\circ}$C) 
& $P_{\rm DOS}$ of Eq. (\ref{P_DOS1}) 
& $P_\rho$ of Eq. (\ref{P_rho2}) & $P_\rho$ (previous values)\\
\hline 
500 
&  0.54 
& 0.73 
& 0.73$\pm$0.02 (Ref. \citen{Goripati}) \\
600 
& 0.62 
& 0.83 
& 0.83$\pm$0.02 (Ref. \citen{Li}) \\
\hline 
\vspace{-0.8cm}
\end{tabular}
\end{center}
\label{table2}
\end{table} 


\begin{thebibliography}{9}
\bibitem{non-HMF}
As seen from Fig. \ref{amr}(d), 
systems with $r_m$$<$1 and $\Delta \rho/\rho$$<$0 can have $x \ll 1$, 
which corresponds to nearly HMFs. 
In contrast, systems with $r_m$$\sim$1 and $\Delta \rho/\rho$$\sim$0 
take a relatively large $x$. 
Only these systems should actually be called non-HMFs.
\bibitem{gamma}
Note that 
$x_s$ in Eq. (\ref{P_rho0}) 
[i.e., $x$ of Eq. (\ref{x_pm1})] 
contains not $\gamma$ but $|\Delta \rho/\rho|/c$ 
with 0$<$$|\Delta \rho/\rho|/c$$<$1.
\bibitem{thermal}
Note that 
real HMFs may exhibit $P_\rho$$\lesssim$1 at finite $T$ 
because the down-spin electrons 
are thermally excited from the valence band to the conduction band. 
Therefore, in the HMF case of $x$=0, 
Eq. (\ref{P_rho1}) is valid at 0 K and 
appropriate for 
$k_{\mbox{\tiny B}}T$$\ll$$E_c-E_{\mbox{\tiny F}}$ 
and $k_{\mbox{\tiny B}}T$$\ll$$E_{\mbox{\tiny F}}-E_v$.
\bibitem{comment}
We utilize the same $y_\uparrow$ as that of the fcc Ni 
because of the same crystal structure and 
the small difference in the number of electrons.
\bibitem{condition}
The CFGG with the $L$2$_1$ structure 
has 
$E_c - E_{\mbox{\tiny F}}$$\sim$0.5 eV and 
$E_{\mbox{\tiny F}}-E_v$$\sim$0.2 eV,\cite{BSDChS} 
which 
are much larger than $k_{\mbox{\tiny B}}T$=0.026 eV, 
with $T$=300 K.
\bibitem{r_m<1}
In this study, we consider the case of $r_m$$\le$1 
because $P_{\rm DOS}$ with $r_m$$>$1 tends to be very small. 
Although $P_\rho$ with $r_m$=1.15 agrees with the previous value at 
$T_{\rm ann}$=600 $^{\circ}$C, 
$P_{\rm DOS}$ with $r_m$=1.15 becomes 0.42, 
which is less than $|P_{\rm DOS}|$ ($\sim$0.5) of Co. 
This $P_{\rm DOS}$ (=0.42) appears to be inappropriate for 
the present CFGG,\cite{Sakuraba} which is regarded as the nearly HMF.
\bibitem{r_m}
As a different method from the present one, 
we note that 
$r_m$ may be roughly evaluated from the above-mentioned equation, 
$m_\sigma$=$\hbar^2 (d^2 E_\sigma /d k_\sigma^2)^{-1}$, 
where 
$E_\sigma$ is obtained by, for example, first-principles calculation.
\end{thebibliography}
\end{document}